\documentclass{PoS}

\ShortTitle{Renormalization of resonance propagator}

\abstract{We study various
aspects of the renormalization of the Resonance Chiral Theory at the
one-loop level using a spin-one resonance propagator as a concrete
example. We calculate explicitly the one-loop
self-energy within the antisymmetric tensor field formalism,
briefly discuss the general structure of the corresponding
propagator obtained by means of the Dyson re-summation and give a
classification of the propagating degrees of freedom. We find that
additional pathological poles (negative norm ghosts or tachyons)
are unavoidably generated and various scenarios according to their
position are possible. We also briefly comment on the eventual
dynamical generation of the opposite parity resonances which are
frozen at the tree level and discuss the role of appropriate
symmetry which could prevent such a scenario.}
\FullConference{International Workshop on Effective Field Theories: from the pion to the upsilon \\
        February 2-6 2009\\
        Valencia, Spain}

\title{Note on renormalization of the spin-1 resonance propagator at one loop order}
\author{Karol Kampf$^{\,ab}$, Ji\v r\'i Novotn\'y$^{\,b}$ and
\speaker{Jaroslav Trnka}$^{\,bc}$ \\
$^a$ Paul Scherrer Institut, Ch-5232 Villigen PSI, Switzerland\\
$^b$ Charles University, Faculty of Mathematics and Physics, IPNP, V Hole\v{s}ovi\v{c}k\'ach 2, CZ-180 00 Prague, Czech Republic\\
$^c$ Department of Physics, Princeton University, 08540 Princeton, New
Jersey, USA\\
E-mail: jtrnka@princeton.edu}

\renewcommand{\logo}
{\parbox{\textwidth}{\begin{flushright} PUPT-2299 \\ PSI-PR-09-06
\end{flushright}\vspace*{-0.3cm}
\includegraphics{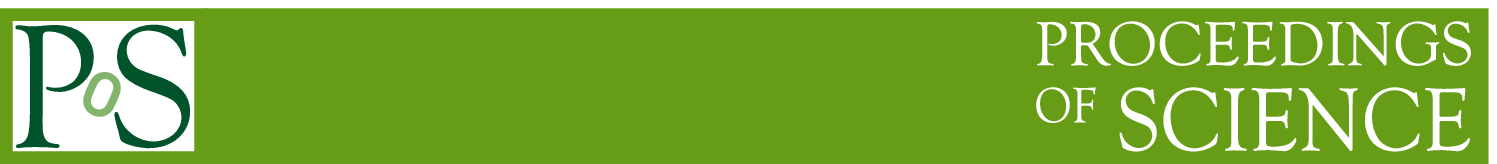}}
}

\begin{document}

\section{Introduction}

Recently, effective theories have become a very efficient tool
in particle physics. As far as the strong interactions are
concerned, the Chiral
perturbation theory ($\chi PT$) \cite{Weinberg}, \cite{Gasser1}, \cite%
{Gasser2} as a low energy effective theory of QCD provides us with a
rigorously defined simultaneous double expansion of the Green functions of
the quark currents in powers (and logarithms) of external momenta and quark
masses which is valid in the energy range $E\ll \Lambda _{H}\sim 1$ GeV.
The scale $\Lambda _{H}$ corresponds to the mass of the lowest resonances
which are separated from the relevant low energy QCD degrees of freedom,
namely the (pseudo)Goldstone bosons (PGB) which are identified with the
members of the lightest pseudoscalar octet, by a mass gap. Thanks to this
mass gap the dynamics of the higher energy degrees of freedom $E\gtrsim
\Lambda _{H}$ can be taken into account effectively and parametrized by
means of the low energy constants (LEC). The corresponding low energy
expansion is therefore possible and well behaved. The formal structure and
the technical aspects of $\chi PT$ are perfectly understood and the recent
calculations reached the two-loop level which corresponds to the order $%
O(p^{6})$ within the chiral power counting \cite{Bijnens:2006zp}.

However, an extension of this successful method to the intermediate energy region $%
\Lambda _{H}\leq E<2$ GeV is more problematic. The set of
relevant degrees of freedom enlarges and contains not only the PGB but
also the low lying resonances. These are not separated by a mass gap
from the
rest of the spectrum and therefore the formal expansion in the spirit of $%
\chi PT$ (\emph{i.e.} a simultaneous expansion in the momenta and
both quark and resonance masses) cannot be expected to be
well-founded. Fortunately, another type of effective Lagrangian
description of this region exists which is based on the large
$N_{C}$ expansion as well as on the high energy constraints derived
from the operator product expansion (OPE). This was introduced in the pioneering works \cite{Ecker1},
\cite{Ecker2} and now it is known as as a Resonance Chiral Theory
($R\chi T$). It becomes extremely useful for the estimates of the
LEC in terms of the resonance parameters \cite{Ecker1},
\cite{Kampf:2006bn}, \cite{Cirigliano:2006hb}, which is necessary
in order to connect the recent $O(p^{6})$ predictions of $\chi PT$
with physical data (cf. also \cite{BJhere}). The theory is organized according to the large
$N_{C}$ expansion: the interaction vertices are accompanied with the
appropriate power of $1/\sqrt{N_{C}}$ for each meson field, the
leading order contributions to the Green function are given by the tree graphs and
each loop brings about one additional power of $1/N_{C}$. Taking
just one resonance multiplet for each channel and matching such a
truncated theory in the UV region with the OPE (which corresponds to
the so-called Minimal Hadronic Ansatz) and with $\chi PT$ in the
infrared was proved to be sufficient to saturate the values of the
$O(p^{4})$ LEC successfully at the leading order in $1/N_{C}$
\cite{Ecker1}.

Such a leading order matching suffers from the fact that the LEC depend
on the renormalization scale. Therefore, this scale has to be fixed
at some value (the saturation scale) at which the renormalization
scale independent results of tree level $R\chi T$ are sewed with
$\chi PT$. This is one of the reasons why to go beyond the leading
order and match $\chi PT$ with the one-loop $R\chi T$. Also from the
phenomenological point of view, the loops are inevitable in order to
preserve (perturbative) unitarity and to generate finite resonance
widths. It is therefore desirable to investigate the one-loop $R\chi
T$ in more details \cite{vecform}.

However, because the Weinberg formula \cite{Weinberg}
(according to which the loop calculations are organized within $\chi
PT$) cannot be straightforwardly generalized to the case of $R\chi
T$, new aspects of the renormalization procedure are expected.
Namely, because of the presence of a new scale corresponding to the
mass of the resonances and as a result of the nontrivial structure
of the higher-spin resonance propagators, we can encounter mixing of
the usual chiral orders in the process of the renormalization. Also,
higher than expected chiral order of counterterms might be necessary
already at the one-loop level. Furthermore, because the spin-one
particles are described using fields transforming under reducible
representation of the rotation group, new degrees of freedom (which
were frozen at the tree-level) can come back to the game due to the loop
corrections.

In this paper we would like to concentrate on a particular example
of the renormalization of the one-loop spin-one resonance
self-energy and the construction of the corresponding resonance
propagator using a concrete interaction Lagrangian. We will use the
antisymmetric tensor field for definiteness, since in such a formalism all
the above aspects of the renormalization procedure can be
illustrated. In addition we will briefly discuss the problems
connected with the appearance of additional poles in the propagator
obtained by means of Dyson re-summation of the one-particle
irreducible insertions. Because the one loop corrections to the
self-energy might be relatively large, these additional poles might
lie near the region for which we assume $R\chi T$ to be valid.
Moreover, the self-energy has higher order growth in the UV region
than usual and therefore some of these poles could be negative norm
ghosts or tachyons \cite{Slovak}. This might introduce well known
problems with the physical interpretation of the theory due to the
violation of unitarity or causality. Due to the lack of appropriate
symmetry, some of the poles correspond to one particle states with
opposite parity than the original degrees of freedom of the tree
Lagrangian. We will also briefly discuss the possibility to
interpret the non-pathological poles as dynamically generated higher
resonances, as was done in \cite{Mont}.

\section{The Lagrangian of Resonance Chiral Theory}

We are going to work in the framework of Chiral perturbation theory
where the Lagrangian is formulated in terms of external sources and
pseudoscalar mesons. They transform as an octet under the group
$SU(3)_V$. We define the chiral building block
\begin{equation}
u(\phi) = \exp\left(i\frac{\phi}{\sqrt2F_0}\right)\,,
\end{equation}
where $\phi=\phi^aT^a$ with $T^a=\lambda^a/\sqrt2$ and
\begin{equation}
\phi(x)=\frac{1}{\sqrt2} \left(
\begin{array}{ccc}
\pi^0+\frac{1}{\sqrt3}\eta & \sqrt2\pi^+ & \sqrt2K^+ \\
\sqrt2\pi^- & -\pi^0+\frac{1}{\sqrt3}\eta & \sqrt2K^0 \\
\sqrt2K^- & \sqrt2\overline{K}^0 & -\frac{2}{\sqrt3}\eta
\end{array}
\right)
\end{equation}
is the matrix describing the pseudoscalar mesons fields. The Goldstone
bosons are parametrized by the elements $u(\phi)$ of the coset space $%
SU(3)_L\times SU(3)_R/SU(3)_V$, transforming as
\begin{equation}
u(\phi) \mapsto V_R u(\phi) h(g,\phi)^{-1} = h(g,\phi)u(\phi)V_L^{-1}
\end{equation}
under a general chiral rotation $g=(V_L,V_R)\in G$ in terms of the $%
SU(3)_V$ compensator field $h(g,\phi)$.

The Resonance Chiral Theory enlarges the number of degrees of freedom of $\chi PT$
by including also massive multiplets. Let us now restrict
ourselves to the octet of vector resonances $1^{--}$ which is the
subject of our interest. There are several possibilities how to
choose corresponding interpolating fields for them \cite{Ecker1,
Ecker2,Kampf:2006yf}. In this article we use the antisymmetric
tensor field which can be written as
\begin{equation}
V_{\mu \nu }=\left(
\begin{array}{ccc}
\frac{1}{\sqrt{2}}\rho ^{0}+\frac{1}{\sqrt{6}}\omega _{8}+\frac{1}{\sqrt{3}}%
\omega _{0} & \rho ^{+} & K^{\ast +} \\
\rho ^{-} & -\frac{1}{\sqrt{2}}\rho ^{0}+\frac{1}{\sqrt{6}}\omega _{8}+\frac{%
1}{\sqrt{3}}\omega _{3} & K^{\ast 0} \\
K^{\ast -} & \overline{K}^{\ast 0} & -\frac{2}{\sqrt{6}}\omega _{8}+\frac{1}{%
\sqrt{3}}\omega _{0}
\end{array}%
\right) _{\mu \nu }\,.
\end{equation}%
The fields $V_{\mu \nu }$ transform in the nonlinear realization of the $%
U(3)_{L}\times U(3)_{R}$ according to the prescription%
\[
V_{\mu \nu }\mapsto h(g,\phi )V_{\mu \nu }h(g,\phi )^{-1}.
\]%
The Lagrangian for these field is then
\begin{equation}
\mathcal{L}_{V}=-\frac{1}{2}\langle \nabla _{\mu }V^{\mu \nu }\nabla
^{\alpha }V_{\alpha \nu }\rangle +\frac{1}{4}M^{2}\langle V_{\mu \nu }V^{\mu
\nu }\rangle +\mathcal{L}_{int}\,,  \label{Lagr}
\end{equation}%
where the relevant interaction part (contributing to the
renormalization of the resonance self-energy which is of our
interest) is
\begin{eqnarray}
\mathcal{L}_{int} &=&\frac{\mathrm{i}G_{V}}{2\sqrt{2}}\langle V^{\mu \nu
}[u_{\mu },u_{\nu }]\rangle +d_{1}\epsilon _{\mu \nu \alpha \sigma }\langle
\{V^{\mu \nu },V^{\alpha \beta }\}\nabla _{\beta }u^{\sigma }\rangle
\nonumber \\
&&+d_{3}\epsilon _{\alpha \beta \mu \lambda }\langle \{\nabla _{\nu }V^{\mu
\nu },V^{\alpha \beta }\}u^{\lambda }\rangle +d_{4}\epsilon _{\rho \sigma
\mu \alpha }\langle \{\nabla ^{\alpha }V^{\mu \nu },V^{\rho \sigma }\}u_{\nu
}\rangle +\dots  \label{Int}
\end{eqnarray}

\section{Structure of poles}

Let us briefly recall the basic properties of the Lagrangian for
spin-1 fields and of the corresponding propagator within the
antisymmetric tensor field formalism \cite{Ecker1}. We start with
the most general free Lagrangian
\begin{equation}
\mathcal{L}_{V}=\frac{\alpha }{2}\langle \partial _{\mu }V^{\mu \nu
}\partial ^{\alpha }V_{\alpha \nu }\rangle +\frac{\beta }{4}\langle \partial
_{\alpha }V^{\mu \nu }\partial ^{\alpha }V_{\mu \nu }\rangle +\frac{1}{4}%
M^{2}\langle V_{\mu \nu }V^{\mu \nu }\rangle\,,  \label{Lagr2}
\end{equation}%
which leads to the propagator
\begin{equation}
\Delta _{\mu \nu \rho \sigma }^{V}(p)=\frac{2}{(\alpha +\beta )p^{2}+M^{2}}%
\Pi _{\mu \nu \rho \sigma }^{L}+\frac{2}{\beta p^{2}+M^{2}}\Pi _{\mu \nu
\rho \sigma }^{T}\,,
\end{equation}%
where $\Pi ^{T}$ and $\Pi ^{L}$ are projectors
\begin{eqnarray}
\Pi _{\mu \nu \alpha \beta }^{T} &=&\frac{1}{2}(g_{\mu \alpha }g_{\nu \beta
}-g_{\mu \beta }g_{\nu \alpha })-\frac{1}{2p^{2}}\left( g_{\mu \alpha
}p_{\nu }p_{\beta }-g_{\mu \beta }p_{\nu }p_{\alpha }+g_{\nu \beta }p_{\mu
}p_{\alpha }-g_{\nu \alpha }p_{\mu }p_{\beta }\right)\,,  \nonumber \\
\Pi _{\mu \nu \alpha \beta }^{L} &=&\frac{1}{2p^{2}}\left( g_{\mu \alpha
}p_{\nu }p_{\beta }-g_{\mu \beta }p_{\nu }p_{\alpha }+g_{\nu \beta }p_{\mu
}p_{\alpha }-g_{\nu \alpha }p_{\mu }p_{\beta }\right)\,.
\end{eqnarray}%
The propagator $\Delta _{\mu \nu \rho \sigma }^{V}(p)$ has two poles in general.
In order to obtain just one pole (and assuming $M^{2}>0$)
we have to fix $\alpha =-1$, $\beta =0$ (which is in agreement with
(\ref{Lagr}), see \cite{Ecker1} for further details) that leads to
the Lagrangian
\begin{equation}
\mathcal{L}=-\frac{1}{2}\langle \partial _{\mu }V^{\mu \nu }\partial
^{\alpha }V_{\alpha \nu }\rangle +\frac{1}{4}M^{2}\langle V_{\mu \nu }V^{\mu
\nu }\rangle\,.  \label{Lagr3}
\end{equation}%
This Lagrangian is generally used for the description of any spin-1
resonances in $R\chi T$. From (\ref{Lagr3}) we get the usual
propagator
\begin{equation}
\Delta _{\mu \nu \rho \sigma }^{V}(p)=-\frac{2}{p^{2}-M^{2}}\Pi _{\mu \nu
\rho \sigma }^{L}+\frac{2}{M^{2}}\Pi _{\mu \nu \rho \sigma }^{T}\,.
\end{equation}
Provided that under parity and charge conjugation $V_{\mu \nu
}\mapsto V^{\mu \nu }$and $V_{\mu \nu }\mapsto -V_{\mu \nu }^{T}$
respectively, the pole in $\Pi ^{L}$ sector corresponds to a
$1^{--}$ resonance . Let us show that the possible pole of $\Delta
_{\mu \nu \rho \sigma }^{V}(p)$ in the $\Pi^{T}$ sector corresponds to an
opposite parity $1^{+-}$ resonance.

The case of $1^{+-}$ resonances was studied in detail in
\cite{Ecker:2007us}.
Starting with the field $B_{\mu \nu }$ describing a $1^{+-}$ resonance (now $%
B_{\mu \nu }\mapsto -B^{\mu \nu }$ under parity) we can write the
same Lagrangian (\ref{Lagr3}) (just replacing $V_{\mu \nu
}\rightarrow B_{\mu \nu }$). Now, we can introduce a field $U_{\mu
\nu }=\frac{1}{2}\epsilon _{\mu \nu \alpha \beta }B^{\alpha \beta
}$ which has the same transformation properties with
respect to parity as the field $V_{\mu \nu}$. Rewriting the Lagrangian (\ref%
{Lagr3}) in terms of $U_{\mu \nu }$ we find
\begin{equation}
\mathcal{L}=\frac{1}{12}\langle H_{\alpha \mu \nu }H^{\alpha \mu \nu
}\rangle -\frac{1}{4}M^{2}\langle U_{\mu \nu }U^{\mu \nu }\rangle\,, \qquad %
\mbox{where}\qquad H_{\alpha \mu \nu }=\partial _{\lbrack \alpha }U_{\mu \nu
]_{cycl}}
\end{equation}%
and the corresponding propagator is
\begin{equation}
\Delta _{\mu \nu \rho \sigma }^{U}(p)=-\frac{2}{M^{2}}\Pi _{\mu \nu \rho
\sigma }^{L}+\frac{2}{p^{2}-M^{2}}\Pi _{\mu \nu \rho \sigma }^{T}\,.
\end{equation}%
Because the $U_{\mu \nu }$ fields describe $1^{+-}$ resonances and have
the same quantum numbers as $V_{\mu \nu }$, the poles in the $\Pi ^{T}$
sector found in the propagator for $V_{\mu \nu }$ indicate the presence
of $1^{+-}$ resonances.
For the free field case there is nothing like that because we fixed $\alpha $, $%
\beta $ to have just one pole in the $\Pi ^{L}$ sector. If we take
$\beta \neq 0$ in
(\ref{Lagr2}) then the additional pole (in the $\Pi ^{T}$ sector) is a ghost (for $%
\beta \,<0$) or a tachyon (for $\beta >0$). Therefore, we can not have
both types of poles at the tree level.

\section{Renormalization of the propagator}

In the general case when the loop corrections are taken
into account one obtains
\begin{equation}
\Delta _{\mu \nu \rho \sigma }^{V}(p)=-\frac{2}{p^{2}-M^{2}-\Sigma
_{L}(p^{2})}\Pi _{\mu \nu \rho \sigma }^{L}+\frac{2}{M^{2}+\Sigma _{T}(p^{2})%
}\Pi _{\mu \nu \rho \sigma }^{T}\,,
\end{equation}%
where $\Sigma ^{L}(p^{2})$ and $\Sigma ^{T}(p^{2})$ are
the self-energies which are determined at the one-loop level by the
following Feynman graphs:
\begin{center}
\includegraphics[width=8cm,angle=0]{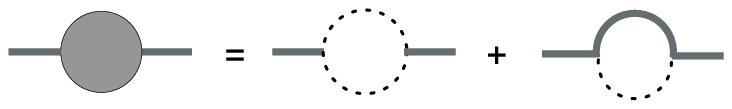}
\end{center}
where the thick lines denote resonances and dashed lines Goldstone
bosons. We use the interaction Lagrangian (\ref{Int}) that keeps
just leading order operators (in the number of derivatives) with no
more than two resonances. The counterterm Lagrangian (terms with up
to six derivatives are needed) is
\begin{equation}
\mathcal{L}_{ct}=\mathcal{L}_{ct}^{(0)}+\mathcal{L}_{ct}^{(2)}+\mathcal{L}%
_{ct}^{(4)}+\mathcal{L}_{ct}^{(6)}
\end{equation}%
with
\begin{eqnarray}
\mathcal{L}_{ct}^{(0)} &=&\frac{1}{2}M^{2}Z_{M}\langle V^{\mu \nu }V_{\mu
\nu }\rangle ,  \nonumber \\
\mathcal{L}_{ct}^{(2)} &=&\frac{1}{2}Z_{R}\langle \nabla _{\alpha }V^{\alpha
\mu }\nabla ^{\beta }V_{\beta \mu }\rangle +\frac{1}{4}Y_{R}\langle \nabla
_{\alpha }V^{\mu \nu }\nabla ^{\alpha }V_{\mu \nu }\rangle ,  \nonumber \\
\mathcal{L}_{ct}^{(4)} &=&\frac{1}{4}X_{R1}\langle \nabla ^{2}V^{\mu \nu
}\{\nabla _{\nu },\nabla ^{\sigma }\}V_{\beta \mu }\rangle +\frac{1}{8}%
X_{R2}\langle \{\nabla _{\nu },\nabla _{\alpha }\}V^{\mu \nu }\{\nabla
^{\sigma },\nabla ^{\alpha }\}V_{\mu \sigma }\rangle +\dots\,,  \label{Lct}
\end{eqnarray}%
where we do not write explicitly all dimension four and six
operators. For the self-energies we find
\begin{eqnarray}
\Sigma _{L}(x) &=&M^{2}\left( \frac{M}{4\pi F}\right) ^{2}\left[
\sum_{i=0}^{3}{\alpha }_{i}x^{i}-\left( \frac{1}{2}\left( \frac{G_{V}}{F}%
\right) ^{2}x^{2}\widehat{B}(x)+\frac{40}{9}d_{3}^{2}(x^{2}-1)^{2}\widehat{J}%
(x)\right) \right]\,,  \nonumber \\
\Sigma _{T}(x) &=&M^{2}\left( \frac{M}{4\pi F}\right) ^{2}\left[
\sum_{i=0}^{3}{\beta }_{i}x^{i}+\frac{20}{9}\left(
2d_{3}^{2}+(d_{3}^{2}+6d_{3}d_{4}+d_{4}^{2})x+2d_{4}^{2}x^{2}\right)
(x-1)^{2}\widehat{J}(x)\right],
\end{eqnarray}%
where $x=p^{2}/M^{2}$ and $\hat{B}(x)$, $\hat{J}(x)$ are loop functions:
\begin{equation}
\hat{B}(x)=1-\ln (-x),\qquad \qquad \hat{J}(x)=\frac{1}{x}\left[ 1-\left( 1-%
\frac{1}{x}\right) \ln (1-x)\right]
\end{equation}%
and $\alpha _{i}$, $\beta _{i}$ are renormalization scale independent
combinations of the couplings and logs, \emph{e.g.}
\begin{equation}
{\alpha }_{0}=\left( \frac{4\pi F}{M}\right) ^{2}Z_{M}^{r}({\mu })-\frac{40}{%
3}d_{1}^{2}\ln \frac{M^{2}}{{\mu }^{2}}-\frac{20}{9}(3d_{1}^{2}-d_{3}^{2})\,.
\end{equation}%
The complete result can be found in \cite{clanek}.

We see that $\Sigma _{T}(x)$ has generally non-trivial $x$ dependence,
therefore, we can expect the possible presence of poles also in the $\Pi ^{T}$
sector.

As was indicated in \cite{Mont} the spectrum of the propagator poles
is very diverse. One of them can be arranged to correspond to the
original $1^{--}$ resonance we have started with. However, it can be
shown \cite{clanek} that (provided we fix the coupling $d_{3}$
according to the OPE for the $VVP$ correlator \cite{VVP}) there exists a
nonzero minimal number of additional poles in both sectors
irrespective of the actual values of the other couplings in the
interaction and the counterterm Lagrangians (\ref{Lagr}) and
(\ref{Lct}). For general values of resonance couplings we could
obtain bound states, virtual states or resonances and also at least
one of the pathological poles like ghosts or tachyons in both the $\Pi
^{T}$ and the $\Pi ^{L}$ sectors. These additional poles decouple in the
$N_{C}\rightarrow \infty$ limit when the interaction is switched
off, however for actual values of the couplings they might lie near
or even inside the region where we expected originally the validity
of $R\chi T$. We can then assume several possible scenarios.

In the most optimistic one, all the additional poles are far
enough and we can treat them as harmless. Then the theory
effectively (\emph{i.e}. when we consider it in the energy region of
its validity) describes the same number of degrees of freedom as
we started with on the tree level.

Within another possible scenario only the pathologies are situated
far away from the range of the assumed validity of $R\chi T$ (this
condition is tricky to satisfy). Then the non-pathological poles
can be treated as a prediction of the theory and identified as
dynamically generated higher resonance states in both $1^{--}$ and
$1^{+-}$ channels. This mechanism is the same as used in
\cite{Penington}, where the scalar resonances were identified as
the poles of the propagator (dressed with the pseudoscalar loops)
of the bare quark-antiquark "seed".

The worst variant arises when some of the pathological poles
appear within the region of assumed applicability of $R\chi T$; in such a
case the theory will suffer from inconsistences like the loss of unitarity
or acausality.

Let us add several brief remarks concerning the second scenario.
Suppose \emph{e.g.} that $1^{+-}$ resonances are really
generated. The question then is which processes the $\Pi^T$ sector of the propagator can  really affect.
We can easily find that in the most common cases of $VV$ correlator,
pion-vector formfactor or $\pi $-$\pi $ scattering it completely
decouples. However, for other processes like $\rho \rightarrow \pi
^{+}\pi ^{-}\gamma$  or $\pi \gamma -\pi \gamma$ we could obtain
some nonzero contribution from these dynamically generated
resonances due to the Feynman graphs

\begin{minipage}[t]{0.47\textwidth}
\begin{center}
  \includegraphics[width=3.3cm,angle=0]{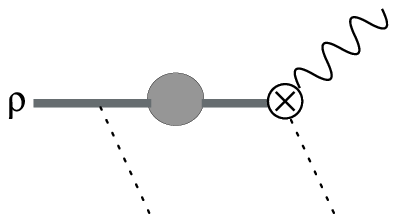}\\
\end{center}
\end{minipage}\hfill
\begin{minipage}[t]{0.47\textwidth}
\begin{center}
  \includegraphics[width=3.7cm,angle=0]{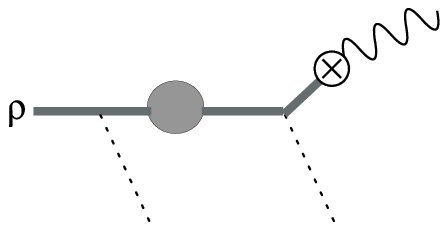}\\
\end{center}
\end{minipage}
and (the thick lines denote resonances, dashed lines Goldstone
bosons and wavy lines ending with the cross vertex indicate the
insertion of the QED current)

\begin{minipage}[t]{0.33\textwidth}
\begin{center}
  \includegraphics[width=4.0cm,angle=0]{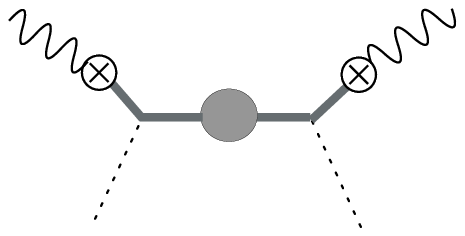}\\
\end{center}
\end{minipage}
\begin{minipage}[t]{0.33\textwidth}
\begin{center}
  \includegraphics[width=3.8cm,angle=0]{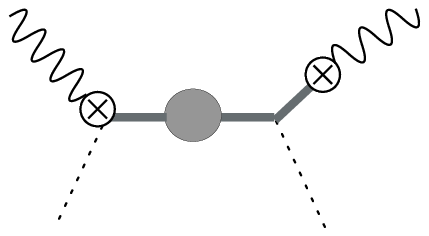}\\
\end{center}
\end{minipage}
\begin{minipage}[t]{0.33\textwidth}
\begin{center}
  \includegraphics[width=3.6cm,angle=0]{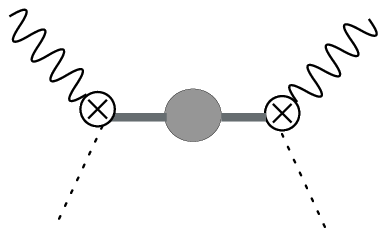}\\
\end{center}
\end{minipage}
respectively.
 Generally, if the vertices in the Lagrangian
that couples to the propagator are invariant under the
transformation
\begin{equation}
V^{\mu \nu }\rightarrow V^{\mu \nu }+\epsilon ^{\mu \nu \alpha \beta
}\partial _{\alpha }\lambda _{\beta }  \label{sym}
\end{equation}%
then the $\Pi ^{T}$ sector of the propagator does not affect a given
process. This transformation corresponds to an accidental symmetry
that some of the vertices posses.

We can also invert our point of view. In the Lagrangian (\ref{Lagr}) we
started with the description of $1^{--}$ resonance and after
renormalization we can get also dynamically generated $1^{+-}$ ones.
So, for this reason or another one we can ask the question: Which
symmetry does prevent the dynamical generation of the $1^{+-}$ at the
one-loop level, \emph{i.e.} when is $\Sigma ^{T}(p^{2})=0$? As an
answer we obtain the same  symmetry as above. If the operators
contributing to the one loop renormalization of the propagator are invariant under
(\ref{sym}) then $\Sigma ^{T}(p^{2})=0$ and there are no $1^{+-}$
resonances. This can be easily seen in the path integral formalism
\cite{clanek}. The price for this is that we have to throw away many
terms in the interaction Lagrangian and finally we may loose the chiral
symmetry. Note also that having the symmetry (\ref{sym}) or any other
mechanism (which would preserve in an ideal case the chiral symmetry)
that freezes the $1^{+-}$ channel, one has to still care about the self-consistency for the $1^{--}$
degree of freedom, \emph{i.e.} one has to still face the three above-mentioned scenarios.

\section{Conclusions}
We have presented here the results of the calculation of the spin-1
resonance self-energy within Resonance Chiral Theory in the
antisymmetric tensor formalism at one-loop. We have found that
additional poles appear in the corresponding Dyson re-summed
propagator some of which are pathological. We have also briefly
discussed various scenarios for their position and consequences
for the physical interpretation of the theory.

\section*{Acknowledgements}

This work was supported in part by the Center for Particle Physics
(project no. LC 527), GACR (project no. 202/07/P249) and by the EU Contract No. MRTN-CT-2006-035482,
\lq\lq FLAVIAnet''. J.~T. is supported by the U.S.Department of
State (International Fulbright S\&T award).

\end{document}